\documentclass{article}
\usepackage[a4paper, margin={1in}]{geometry}
\usepackage[utf8]{inputenc}
\usepackage{amsmath}
\usepackage{graphicx}
\usepackage{enumerate}
\usepackage{natbib}
\usepackage{url} 
\usepackage{tikz} 
\usepackage{accents}
\usepackage{bbm}
\usepackage{bm}
\usepackage{siunitx}
\usepackage{multirow}
\usepackage{setspace}

\newcommand{\Tau}{\mathcal{T}}
\newcommand{\ubar}[1]{\underaccent{\bar}{#1}}
\DeclareMathOperator*{\argmax}{argmax}

\usetikzlibrary {shapes ,arrows , positioning }
\usetikzlibrary{decorations.fractals}
\usetikzlibrary{decorations.pathreplacing}
\usetikzlibrary{arrows.meta}
\usetikzlibrary{automata,chains}

\title{ Bayesian Semiparametric Model for Sequential Treatment Decisions with Informative Timing }
  \author{ \small Arman Oganisian \\
     \small Department of Biostatistics, Brown University \\
     \small Kelly D. Getz  \\
     \small Department of Biostatistics, Epidemiology, and Informatics, \\ 
     \small University of Pennsylvania \\ 
     \small Todd A. Alonzo  \\
     \small Department of Preventive Medicine,  
     \small University of Southern California \\ 
     \small Richard Aplenc  \\
     \small Division of Oncology, Children's Hospital of Philadelphia \\ 
     \small Jason A. Roy \\ 
     \small Department of Biostatistics and Epidemiology, Rutgers University }
\begin{document}
\maketitle

\begin{abstract}
\singlespacing
We develop a Bayesian semi-parametric model for the estimating the impact of dynamic treatment rules on survival among patients diagnosed with pediatric acute myeloid leukemia (AML). The data consist of a subset of patients enrolled in the phase III AAML1031 clinical trial in which patients move through a sequence of four treatment courses. At each course, they undergo treatment that may or may not include anthracyclines (ACT). While ACT is known to be effective at treating AML, it is also cardiotoxic and can lead to early death for some patients. Our task is to estimate the potential survival probability under hypothetical dynamic ACT treatment strategies, but there are several impediments. First, since ACT was not randomized in the trial, its effect on survival is confounded over time. Second, subjects initiate the next course depending on when they recover from the previous course, making timing potentially informative of subsequent treatment and survival. Third, patients may die or drop out before ever completing the full treatment sequence. We develop a generative Bayesian semi-parametric model based on Gamma Process priors to address these complexities. At each treatment course, the model captures subjects' transition to subsequent treatment or death in continuous time under a given rule. A g-computation procedure is used to compute a posterior over potential survival probability that is adjusted for time-varying confounding. Using this approach, we conduct posterior inference for the efficacy of hypothetical treatment rules that dynamically modify ACT based on evolving cardiac function.
\end{abstract}

\newpage
\doublespacing

\section{Introduction}

\label{sec:intro}
Our application involves a subset of data from the AAML1031 phase III clinical trial run by the Children's Oncology Group (COG). Upon trial enrollment, subjects with pediatric acute myeloid leukemia (AML) were randomized to a four-course treatment sequence. At each course, patients were given a standard chemotherapy treatment comprised of multiple agents which may or may not include anthracycline (ACT). In addition to the standard chemotherapy, some patients were also randomized to receive bortezomib. Despite its demonstrated effectiveness in treating AML, ACT agents are known to be cardiotoxic. The impact of ACT on overall survival is complex - for some it could increase survival while for others it may cause cardiotoxicity and early death. This presents a potential risk-benefit tradeoff associated with ACT. Ahead of each course, physicians conducted an echocardiogram to measure patients' left ventricular ejection fraction (EF) - a marker of cardiac function with lower values indicating worse function. Along with other information, EF is used to decide whether to administer ACT or withhold ACT. The precise ways in which EF is used varies, but often involves withholding ACT if decline in EF is below some relative threshold (e.g. a 10\% decline from baseline \citep{Neuendorff2020}) or if EF is below some absolute threshold (e.g. any EF value below 50\%) - but the utility of these thresholds and their effects on survival are still debated. In a previous trial, COG AAML0531, about 12\% of patients experienced cardiotixicity within a year of treatment initiation. In COG AAML1031, the percentage is 22\% which is likely due to more complete monitoring of patients. Additionally, patients who experienced cardiotoxicity were found to have lower overall survival in both the AAML0531 and AAML1031 trials \citep{Getz2019,Getz2019_2}. Thus, there is strong clinical interest in estimating (and optimizing) the impact of ACT treatment on survival. 

Fortunately, a subset of the AAML1031 patients were treated at centers that contribute to a separate Pediatric Health Information System (PHIS) database, allowing us to assess ACT administration via PHIS billing data. While the data are rich, they pose several impediments to estimation. First, ACT administration is not randomized, but modified as needed according to the patient's clinical status at each treatment course. Thus, its effect on survival is confounded by time-varying variables such as EF and other information available just ahead of each treatment decision. Second, while each subject receives at most four treatment courses, the timing of the courses vary by subject. This is partially because AML chemotherapy causes prolonged neutropenia. Subjects initiate the next course after they achieve hemotalogic recovery and are clinically stable. Since patients with longer recovery times may have greater toxicities, the waiting times between treatment courses may be informative of subsequent ACT decision and survival. Third, patients may die or drop out of the study before completing the full sequence so that the total number of  treatment courses varies across patients as well. Naive approaches such as restricting the analysis to only subjects who have survived through the full treatment sequence will generally yield a biased subset (likely systematically healthier) of subjects. Common approaches such as discretizing the time period (e.g. into months) do not respect the continuous-time nature of the decision-making process, often leading to difficulties in dealing with patients who initiate treatments between discrete periods. Standard g-computation approaches \citep{Robins1986} were designed for settings where all subjects receive, say, the second treatment at the same state of their disease process. These are inappropriate in settings where timing may confound subsequent treatment and survival.

To address these concerns, we frame ACT assignment as a dynamic treatment regime in which a rule maps a patient's available history at each course to a treatment decision. Under certain assumptions about the underlying causal structure, we identify the potential survival probability under a given treatment rule in terms of the cause-specific hazards of subsequent treatment and death as well as the distribution of the time-varying confounders (e.g. EF) at each course. It is recognized that even when the causal assumptions hold, inferences about potential survival probability will be biased if modeling assumptions are violated. This motivates a Bayesian semiparametric specification of the cause-specific hazards that is more robust to misspecification. In particular, we specify proportional hazard models in which the baseline hazards follow a Gamma Process prior. Together, these semiparametric cause-specific hazards characterize patients' transition between successive treatments and death in continuous time. Markov Chain Monte Carlo (MCMC) samplers for such models can be difficult to tune due to the large number of parameters present even with just four time periods, so we outline an adaptive sampling scheme for posterior inference. Finally, a g-computation procedure is outlined that simulates from the continuous-time transition process under hypothetical treatment rules and computes the corresponding potential survival probability. The end product is full posterior inference for potential survival rates (and their functionals) under treatment rules of interest which are appropriately adjusted for time-varying confounding and informative waiting times while offering robustness to misspecification.

While causal estimation and optimization with dynamic treatment rules have been discussed for some time \citep{Robins1986,Murphy2003}, Bayesian nonparametric (BNP) estimation strategies in particular have only recently grown in popularity. A key draw of BNP is the ability to incorporate posterior uncertainty about the decision process into a single posterior distribution for the target estimand while leveraging flexible models that impose only weak structure on the data. Moreover, simulations from this posterior can be transformed to allow inference about functionals of interest. We build upon this literature with several distinct contributions. For instance, \cite{Xu2016} proposed a BNP model for sequential AML treatments based on Dependent Dirichlet Processes, however their setting only required adjustment for a single set of baseline confounders. In our setting, we have important time-varying confounders whose evolution must be modeled over time. \cite{Qian2020} and \cite{Hua2021} also develop Bayesian approaches in a setting with random, potentially informative treatment times. Specifically, \cite{Qian2020} consider optimizing a value function based on a proportion outcome, which was not subject to right-censoring. \cite{Hua2021} develop a Bayesian point-process model of a decision process with a survival outcome where all decision points of interest are before the observed time on study and a Weibull proportional hazard submodel is fit for the event time. In contrast, we have a fixed number of decision points of interest, and not all subjects will survive through the full sequence. Moreover, we take a semiparametric approach to modeling the event time via Gamma Process priors. \cite{Murray2017} propose a design of a randomized clinical trial assessing the efficacy of a treatment sequence where the timing of a third-line therapy is random. In our study, ACT administration is not randomized, which necessitates careful thought about potential outcomes and confounding adjustment. 

In the following sections we describe the observed AAML1031 data structure, formalize and identify our target estimand in terms of potential outcomes, and discuss Bayesian inference. As our approach is tailored to deal with the complexities of our data, we will frequently motivate particular features by referencing the data, but a complete data analysis will be reserved for the last section.

\section{Description of AAML1031-PHIS Data}

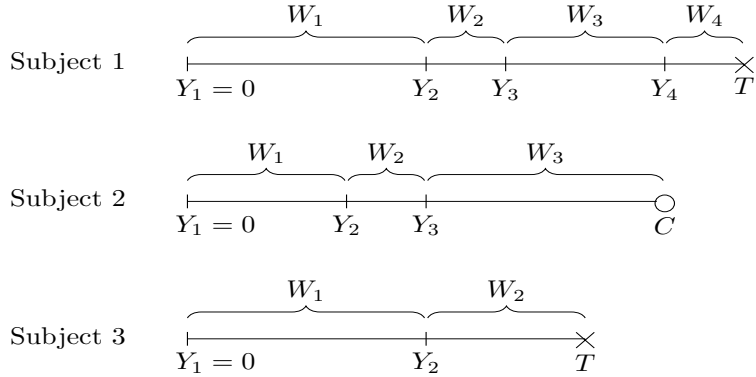
\begin{figure}
\begin{center}
\resizebox{4in}{2in}{%
\begin{tikzpicture}

\draw (-1.5,.5) node[below, yshift=-.6em]{Subject 1};
\draw (0,0pt) -- (7,0pt);
\draw (0,3pt) -- (0,-3pt) node[below, xshift = 10pt]{$Y_1=0$};
\draw (3,3pt) -- (3,-3pt) node[below]{$Y_2$};
\draw (4,3pt) -- (4,-3pt) node[below]{$Y_3$};
\draw (6,3pt) -- (6,-3pt) node[below]{$Y_4$};
\draw[-{Rays[scale=2]} ] (7,3pt) node[below, yshift=-.5em]{$T$};

\draw [decorate,decoration={brace,amplitude=5pt,raise=2ex}]
 (0,0) -- (2.98,0) node[midway,yshift=2em]{$W_1$};
\draw [decorate,decoration={brace,amplitude=5pt,raise=2ex}]
 (3.02,0) -- (3.98,0) node[midway,yshift=2em]{$W_2$};
\draw [decorate,decoration={brace,amplitude=5pt,raise=2ex}]
 (4.02,0) -- (5.98,0) node[midway,yshift=2em]{$W_3$};
\draw [decorate,decoration={brace,amplitude=5pt,raise=2ex}]
 (6.02,0) -- (6.98,0) node[midway,yshift=2em]{$W_4$};

\draw (-1.5,-1.7) node[below]{Subject 2};
\draw (0,-2) -- (5.9,-2);
\draw (0,-2.1) -- (0,-1.9) node[below,xshift = 10pt,yshift=-5pt]{$Y_1=0$};
\draw (2,-2.1) -- (2,-1.9) node[below, yshift=-5pt]{$Y_2$};
\draw (3,-2.1) -- (3,-1.9) node[below, yshift=-5pt]{$Y_3$};
\draw[-{Circle[open, scale=2]} ] (6,-1.9) node[below, yshift=-.6em]{$C$};

\draw [decorate,decoration={brace,amplitude=5pt,raise=2ex}]
 (0,-2) -- (1.98,-2) node[midway,yshift=2em]{$W_1$};
\draw [decorate,decoration={brace,amplitude=5pt,raise=2ex}]
 (2.02,-2) -- (2.98,-2) node[midway,yshift=2em]{$W_2$};
\draw [decorate,decoration={brace,amplitude=5pt,raise=2ex}]
 (3.02,-2) -- (5.98,-2) node[midway,yshift=2em]{$W_3$};

 \draw (-1.5,-3.7) node[below]{Subject 3};
\draw (0,-4) -- (5,-4);
\draw (0,-4.1) -- (0,-3.9) node[below, xshift=10pt, yshift=-5pt]{$Y_1=0$};
\draw (3,-4.1) -- (3,-3.9) node[below, yshift=-5pt]{$Y_2$};
\draw[-{Rays[scale=2]} ] (5,-3.9) node[below, yshift=-.6em]{$T$};

\draw [decorate,decoration={brace,amplitude=5pt,raise=2ex}]
 (0,-4) -- (2.98,-4) node[midway,yshift=2em]{$W_1$};
\draw [decorate,decoration={brace,amplitude=5pt,raise=2ex}]
 (3.02,-4) -- (5,-4) node[midway,yshift=2em]{$W_2$};

\end{tikzpicture}
}
\end{center}
\caption{\small Some possible patient trajectories in our data. While at most $K=4$ treatments are possible, the timing of the $k^{th}$ treatment, $Y_k$, varies. Patients may die after completing the full sequence (Subject 1) but may also drop out (Subject 2) or die (Subject 3) before completing the full sequence. \label{fig:diag}}
\end{figure}

Consider a scenario where patients move through a sequence of $K$ treatment courses, indexed by $k=1, 2, \dots, K$. In our AML study, $K=4$. At the beginning of course $k$, physicians monitor a vector of $P$ patient characteristics, $L_k \in \mathcal{L}_k $, to guide their decision to include ACT or withhold ACT in favor of treating with non-anthracyclines (nACT) . We let $A_k \in \mathcal{A}_{k}=\{0,1\}$ denote this decision with $A_k=1$ indicating ACT inclusion. In our analysis, these include time-constant features such as sex and genotype information as well as time-varying features like ejection fraction and presence bloodstream infection. These features typically affect both treatment decisions and survival, making them important time-varying confounders which must be adjusted for in order to estimate the impact of the treatment sequence on survival. Importantly, while the maximum number of treatment courses is fixed at $K$, patients initiate each course at different times depending on how quickly they recover from their previous course. Treatment course $k$ is initiated at time $Y_k>0$ with available information $H_k = ( \bar L_k,  \bar A_{k-1}, \bar Y_{k} ) \in \mathcal{H}_k$, where the overbar $\bar X_k = (X_1, X_2, \dots, X_k) $ denotes the history of $X$ up to and including $X_k$. We use underbar notation to denote future observations, e.g. $\ubar X_k = (X_k, X_{k+1}, \dots, X_K)$. All patients undergo their first treatment course, $A_1$, at time $Y_1:=0$. The remaining treatment times $Y_k$ for $k>1$ vary across subjects and are measured relative to $Y_1=0$. We adopt the convention that at time $Y_k$, $L_k$ is measured before $A_k$ is administered. Finally, let $T>0$ denote the patient's survival time, also measured from $Y_1=0$. In our data, subjects may die before completing the full sequence of $K$ treatments. Additionally, patients may drop out at any point after the first treatment, inducing right-censoring. This means that each subject will undergo a random number of treatment courses, $\kappa \in \{1, 2, \dots, K \}$. Since all subjects undergo the first treatment course, $\kappa\neq0$. Let $W = \min(T, C)$ be the observed time in study, which is the minimum of death time and censoring time, $C>0$, and let $\delta = I(T < C)$ be the corresponding indicator. For course $k=1, \dots, K$ we define a sequence of waiting times until the next event. Specifically, for subjects still in the study at the start of treatment course $k$ (i.e. $\kappa \geq k$),  $W_k$ is the time from treatment $k$ initiation, $Y_{k}$, to either death at time $T$, drop out at time $C$, or subsequent treatment at time $Y_{k+1}$ - whichever comes first. Thus, the observed waiting time is $W_k = \min(Y_{k+1} , T , C ) - Y_{k}$ and the corresponding indicator $\delta_{k} \in\{1, 0, -1\}$ indicates a transition to death, censoring, or subsequent treatment respectively. Figure \ref{fig:diag} depicts a few possible patient trajectories. We additionally define cause-specific transition indicators for death and subsequent treatment, $\delta_{Tk} = I(\delta_{k}=1)$ and $\delta_{Yk}=I(\delta_k = -1)$. At the last treatment course, there is no subsequent treatment so $W_K = \min(T , C ) - Y_{K}$, $\delta_{K} \in\{1, 0\}$, $\delta_{TK} = \delta_K$ and $\delta_{YK}$ is undefined. 

\begin{figure}
\begin{center}
\includegraphics[width=1\linewidth]{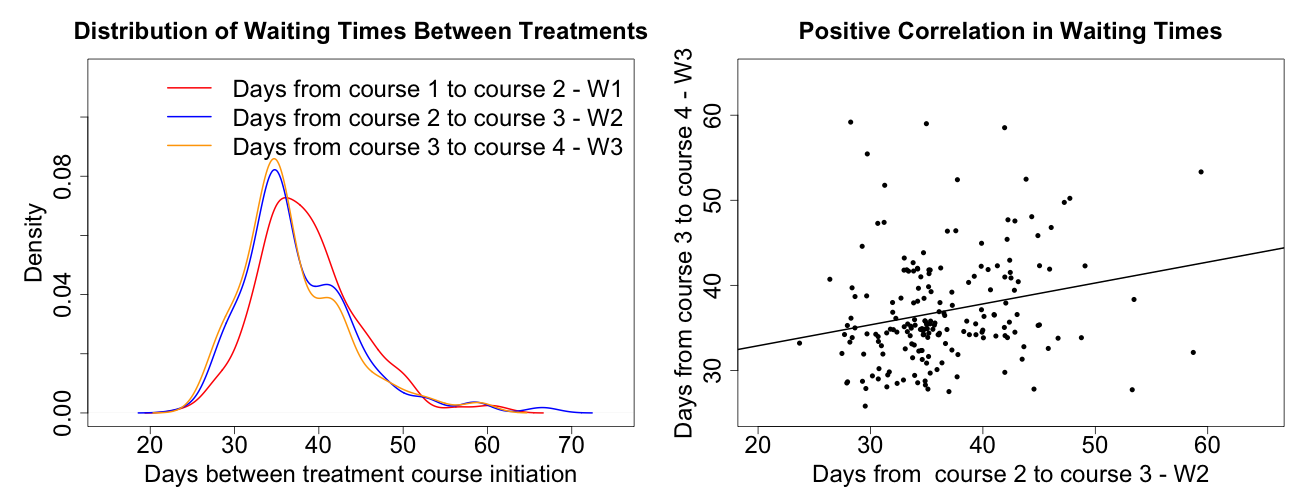}
\end{center}
\caption{ Left: Distribution of waiting times between treatment courses among those at-risk. Note the long tail as some patients take especially long to recover before starting their next course. Right: scatterplot of observed waiting times between course 2 and 3 ($W_2$) and course 4 and 3 ($W_3$) among those who had all four treatment courses visualizing positive association in waiting times. \label{fig:descplots} }
\end{figure}

In this setup, the observed time for a subject who underwent $\kappa$ treatment courses can be expressed as the sum of the waiting times between events $W = \sum_{j=1}^\kappa W_j $. We equivalently write the history available just before treatment decision $A_k$ as $H_k = (\bar L_k, \bar A_{k-1}, \bar Y_k ) =  (\bar L_k, \bar A_{k-1}, \bar W_{k-1})$, since the $ \bar W_{k}$ are just transformations of $\bar Y_k$. We observe data on $i=1, 2, \dots, n$ independent subjects. The data for subject $i$ consists of $\mathcal{D}_i = (\bar L_{\kappa_i}, \bar A_{\kappa_i}, \bar W_{\kappa_i} , \bar \delta_{Y\kappa_i}, \bar \delta_{T\kappa_i})$ and we let $\mathcal{D} = \{ \mathcal{D}_i \}_{i=1}^n$ denote the full data.

The left panel of Figure \ref{fig:descplots} shows appreciable variation in waiting times around 35 days, with a tail of subjects who start their next course later than average. At any given course, the time since previous course potentially informs the treatment decision (e.g., subjects who took longer than usual to recover should perhaps be treated conservatively) and may also be related to survival time accrued afterward. Indeed, the right panel of Figure \ref{fig:descplots} shows that observed waiting times between courses are positively correlated in these data. These complications, along with time-varying confounding of the treatment sequence, motivate the formal causal approach to estimating treatment effects described in the next section.

\section{Target Estimand and Identification}
Here, we define relevant target estimand in terms of potential outcomes had we assigned treatment via some rule and state the main identification result that expresses this estimand in terms of objects that can be estimated from the observed data. Suppose the treatment decision at course $k$ is the output of decision rule $r_k : \mathcal{H}_k \rightarrow \mathcal{S}_k(h_k)$ that maps from the available history to a treatment in the feasible set, $d_k^r \in \mathcal{S}_k(h_k)\subset \mathcal{A}_k$, and denote the collection of rules as $r = (r_1, r_2, \dots, r_K)$. It is necessary to define feasible subsets of $\mathcal{A}_k$ since, depending on a subject's history $h_k$, some options in $\mathcal{A}_k$ may not be possible. For instance, in the AAML1031, all subjects were generally ineligible for ACT at the third course and so $\mathcal{S}_3(h_3) = \{0\} \subset \mathcal{A}_3$ for all $h_3$. 
The ACT/nACT treatment sequence is not set at baseline, but adapts dynamically to each patient's unique disease progression: at time $Y_1$, we first measure $L_1$ then use it to arrive at a decision $d_1^r = r_1(L_1, Y_1)$ and assign $A_1 = d_1^r$. At the second treatment course, at time $Y_2$, we measure $L_2$ the make decision $d_2^r = r_2(\bar L_2, A_1=d_1^r, \bar Y_2,) = r_2(H_2)$ and set $A_2 = d_2^r$. Similarly, at the $k^{th}$ treatment course, we arrive at decision $d_k^r = r_k(H_k)$ and set $A_k = d_k^r$. We let $\bar d_k^r$ denote the decision history up to and including decision $k$. For $k=2, \dots, K$, let $\kappa_k(\bar d_{k-1}^r) \in\{0,1\}$ indicate whether subject would have survived long enough to initiate course $k$, had previous treatment decision been made via rule $r$. By construction, $ \kappa_k(\bar d_{k-1}^r) = 1$ implies $\kappa_j(\bar d_{j-1}^r) = 1$ for all $j< k$ and  $ \kappa_k(\bar d_{k-1}^r) = 0$ implies $\kappa_j(\bar d_{j-1}^r) = 0$ for all $j>k$. Thus, $\kappa(r) = \min_{k} \{ k :  \kappa_k(\bar d_{k-1}^r) = 1 \} $ denotes the potential number of treatments a patient would have received. For $k=2, \dots, K$ we define potential time of treatment $k$ as $Y_k(r) = Y_k( \bar d_{k-1}^r)$ had they been treated according to rule $r$. Since $\kappa_1=1$ and $Y_{1} = 0$ for all subjects, they are not potential outcomes of $r$. We also define potential survival time under rule $r$, $T(r) = T(\bar d^r_{\kappa(r)}) $, and $W_{k}(r) = W_k( \bar d_k^r ) = \min(T(r), Y_{k+1}(r)) - Y_{k}(r)$ the corresponding potential waiting time from treatment $k$ to the first of $Y_{k+1}(r)$ or $T(r)$. Finally, we define the potential confounder value measured at treatment course $k$, $L_k(r) = L_k(\bar d_{k-1}^r)$. 

Our target estimand is the potential probability of surviving past $t$ under rule $r$, $\Psi^r( t ) = P( T(r) > t )$. Note that $\Psi^r( t ) $ is a function of potential outcomes, not observed data. As is typical in causal inference settings, identification assumptions are required to express $\Psi^r( t )$ -  in terms of observed data, $\mathcal{D}$. Throughout, we assume the following assumptions about the causal mechanism hold:

\begin{enumerate}
	\item {\bf{ Sequential conditional exchangeability}}: at the $k^{th}$ treatment course,  
	$$ \ubar W_k(\bar d_k^r ), \ubar \kappa_{k+1}(\bar d_k^r ), \ubar L_{k+1}(\bar d_k^r ) \perp A_k \mid  \bar A_{k-1} = \bar d_{k-1}^r, \bar L_{k}, \bar W_{k-1}, \kappa \geq k $$
Among those who underwent course $k$, treatment $A_k$ is as good as random conditional on available history. I.e. assignment is unrelated to future potential waiting times, confounder trajectories, and number of subsequent treatments. This can be viewed as emulating a sequentially randomized trial.
	\item {\bf{Treatment positivity}}: $P( A_k = a_k \mid h_k , \kappa \geq k) > 0$ for all histories $f(h_k)>0$. This must hold for each feasible treatment option $a_k \in \mathcal{S}_k(h_k)$. This could be violated structurally if, for instance, subset of patients with history $(\bar a_{k-1}, \bar l_k, \bar w_k)$ would always receive one of the two treatments - rendering the causal effect in this subset undefined.
	\item {\bf{Non-informative censoring}}: The cause-specific hazard of being censored $w$ time units after $Y_k$ is given by $\lambda_C(w \mid h_k, \kappa \geq k )$. Note that it does not depend on potential outcomes conditional on available history, $h_k$. This ensures that censoring is conditionally random and rules out situations in which, say, subjects with systematically higher potential survival times are more likely to be censored. Additionally, we require $P( C\geq Y_k + w \mid W_k( \bar d_k^r) = w, h_k, \kappa \geq k ) > 0$. At each course there is a positive probability of reaching the next event at $W_k( \bar d_k^r) = w$. Else there would be some subgroup in terms of $h_1$ for whom we would never observe the full outcome history.
	\item {\bf{SUTVA} }: For uncensored subjects with observed treatment vector $\bar A_k = \bar a_k$, the potential outcomes under $\bar a_k$ are equal to the observed outcomes: $W_k(\bar a_k) = W_k$, $L_k(\bar a_k) = L_k$, and $\kappa_k(\bar a_k) = \kappa_k$. For
\end{enumerate}
As in standard survival settings, the survival curve, $\Psi^r(t)$, is nonparametrically identifiable only up to the maximum observed death time. Under these assumptions, the joint distribution of the potential outcomes can be identified in terms of observed data in settings where both the number of treatments and timing is random \citep{tsiatis2019}. Various quantities of interest can then computed as integrals over this joint. In our case, $\Psi^r( t ) = \int_{t}^\infty f_{T(r)}( w ) dw$. By the total law of probability, this can be expressed as the joint averaged over the potential number of treatments $\Psi^r( t ) = \int_{t}^\infty \sum_{k=1}^K f_{T(r),  \kappa(r)}( w, \kappa(r) = k) dw$. Identification of  $\Psi^r( t )$ is achieved by identifying $f_{T(r), \kappa(r)}( w, \kappa(r) = k)$ for each $k$. We present the result for $k=K$ and provide a derivation in the supplement
\begin{equation} \label{eq:gcomp}
\small
	\begin{split}
		& f_{T(r), \kappa(r)} ( w, \kappa(r) = K)  = \int_{\bar{ \mathcal{L}}_K} \int_{\bar {\mathcal{W}}_{K-1}} \prod_{k=1}^{K-1} \Big\{ f_k(l_k \mid \bar d_{k-1}^r, \bar l_{k-1}, \bar w_{k-1}, \kappa \geq k ) \\
		& \ \ \ \times S_k(w_k \mid \bar d_k^r, \bar l_{k}, \bar w_{k-1},  \kappa \geq k) \lambda_{Yk}(w_k \mid \bar d_k^r, \bar l_k, \bar w_{k-1},  \kappa \geq k ) \Big\} \\
		& \ \ \ \times f_K(l_k \mid \bar d_K^r, \bar l_{K-1}, \bar w_{K-1}, \kappa \geq K ) S_{TK}( w - \sum_{k=1}^{K-1} w_k \mid \bar d_K^r , \bar l_K, \bar w_{K-1}, \kappa \geq K ) \\
		& \ \ \ \times  \lambda_{TK}( w - \sum_{k-1}^{K-1} w_k \mid \bar d_K^r, \bar l_{K}, \bar w_K, \kappa \geq K) \  d \bar l_K \ d \bar w_{K-1}
		\end{split}
\end{equation}
This is a version of the g-formula \citep{Robins1986} where we integrate over the joint distribution of the observed confounders and waiting times under interventions that are dynamically set via rule $r$, $\bar d_K^r$. The process of computing the integrals involved in \eqref{eq:gcomp} is known as g-computation, which we do via Monte Carlo simulation in practice. Above, $f_k$ is the density function of confounders measured at time point $k$ conditional on available history. Since there is no previous history at $k=1$,  $\bar X_{0} = \emptyset$. So, for instance, in the above we have $ f_1(l_1 \mid \bar a_{0}, \bar l_{0}, \bar w_{0} ) =  f_1(l_1)$. For $k=1, 2, \dots, K-1$, $\lambda_{Yk}(w \mid -) $ is the cause-specific hazard of subsequent treatment after treatment $k$
\begin{equation} \label{eq:haz}
	\begin{split}
		\lambda_{Y k}( w \mid - ) = \lim_{dw\rightarrow 0} dw^{-1} P( w \leq W_{k} <  w + dw, \delta_{k} = -1 \mid W_{k} \geq w,  - )
	\end{split}
\end{equation}
For compactness, the conditioning set is abbreviated as ``$-$''. Conditional on not having died or initiated the subsequent treatment by $w$ time-units after treatment $k$, $\lambda_{Y k}$ is the instantaneous probability of being treated again at that time. Similarly, the function $ \lambda_{Tk}(w \mid -)$ for $k=1, 2, \dots, K$ is the cause-specific hazard of death after treatment course $k$ and has the same form as \eqref{eq:haz} except with $\delta_k = 1$. This represents the instantaneous probability of dying $w$ time units after treatment $k$, conditional on not having died or received treatment $k+1$ by that time. The function $S_k(w \mid - \ ) = \exp( - \int_{0}^w\{ \lambda_{Tk}( s \mid - \ ) + \lambda_{Yk} (s \mid - \ ) ds  \} )$ is the probability of not having died or received subsequent treatment within $w$ time-units after treatment $k$ - i.e. the probability of being at-risk for subsequent treatment or death at $w$. Since there are no more treatments after course $K$, $S_{TK}(w \mid - \ ) = \exp(  - \int_{0}^w \lambda_{TK} ( s \mid - ) ds )$ - which does not involve the hazard of subsequent treatment. These functional forms follow from the standard relationships between hazard, density, and survival functions in the competing risks literature \citep{kalbfleisch2011} - where after each treatment, subsequent treatment and death are competing to be the next event. Alternatively,  \eqref{eq:gcomp} can be understood in terms of transition probabilities between states of treatment and death. In \eqref{eq:gcomp}, the product $P(\delta_k = -1 \rightarrow \delta_{k+1} = -1) = S_k(w_k \mid- ) \lambda_{Yk}(w_k \mid -)$ is the conditional probability of transitioning to treatment course $k+1$ $w_k$ time units after course $k$. Similarly, $P(\delta_k = -1 \rightarrow \delta_{k+1} = 1) = S_k(w_k \mid- ) \lambda_{Tk}(w_k \mid -)$ is the conditional probability of transitioning to death $w_k$ time units after course $k$. Thus these cause-specific hazards characterize how subjects transition to either subsequent treatment or death states, in continuous time, under rule $r$. In the following sections, we propose Bayesian semi-parametric models for the unknowns in \eqref{eq:gcomp}, outline an MCMC procedure for posterior sampling, and describe a g-computation procedure for computing the integrals in  \eqref{eq:gcomp}.
\section{Generative Bayesian Semiparametric Models } \label{sc:models}
The quality of our inference about $\Psi^r(t)$ relies on correctly specified models for $f_k$, $\lambda_{Yk}$, and $\lambda_{Tk}$. Misspecification of these nuisances may bias estimates of $\Psi^r(t)$, \textit{even if} the causal identification assumptions hold. To be more robust to misspecification, we propose proportional hazard models with a flexible baseline hazard specification. For the hazard of subsequent treatment after treatment $k$, we set
\begin{equation} \label{eq:gp}
	\begin{split}
		\lambda_{Yk}(w \mid H_k,  \lambda_{0Yk}, \beta_{Yk} ) & = \lambda_{0Yk}(w) \exp\big( H_k' \beta_{Yk} \big) \\
		\Lambda_{0Yk} & \sim GP(\alpha_{Yk} \Lambda_{Yk}^*) \\
		\beta_{Yk} & \sim f_{\beta_{Yk} }(\beta_{Yk} )
	\end{split}
\end{equation}
Here, $\Lambda_{0Yk}(w) = \int_{0}^w \lambda_{0Yk}(u) du $ is the baseline cumulative hazard and $GP(\alpha \Lambda_{Yk}^*)$ denotes the Gamma Process prior \citep{Kalbfleisch1978} over the unknown baseline cumulative hazard. $ f_{\beta_{Yk}}$ is a prior over the multiplicative effects, $\beta_{Yk}$. Throughout we set this to multivariate normal with density $f_{\beta_{Yk}}=N(0, cI)$. Note that on a hazard ratio scale, $c=1$ is fairly informative - putting most prior mass on ratios in $\exp(\pm 3\cdot1)$. The Gamma Process (GP) is a stochastic process that generates random positive, non-decreasing functions. Realizations from the GP in \eqref{eq:gp} are random cumulative hazards centered around a specified prior cumulative hazard $ \Lambda_{Yk}^*$ with dispersion controlled by $\alpha_{Yk}$. The key property of GPs is that if $\Lambda_{0Yk}(w)$ follows a GP, then the hazard rate over any finite interval follows a Gamma distribution that is centered around the hazard rate over that interval given by the prior $\lambda_{Yk}^*$. This motivates a piecewise constant specification of $\lambda_{0Yk}(w)$ over a pre-specified partition of time $u_{Yk1} = 0 < u_{Yk2} < u_{Yk1} < \dots < u_{J_{Yk}} = \max(W_k)$: $\lambda_{0Yk}(w) = \sum_{j=1}^{J_{Yk}-1x} \mathbbm{1}_{[u_j, u_{j+1})} (w) \lambda_{0Yk}^j$. The Gamma Process induces independent Gamma priors on the $j^{th}$ hazard rates $(u_{j+1} - u_j)\lambda_{0Yk}^j \sim Gam\big( \alpha_{Yk} ( \Lambda_{Yk}^*(u_{j+1}) -  \Lambda_{Yk}^*(u_j)  ), \alpha_{Yk} \big)$. A finer partition leads to a higher-dimensional, flexible hazard model. On the other extreme, a single partition completely shrinks to a one-dimensional constant hazard model. With a fine partition, this provides an alternative to, say, a $Weibul(a, s)$ that imposes a rigid form for the hazard, $\lambda_{0Yk}(w) =  \frac{a}{s^a} w^{a-1}$. We take an empirical approach that sets $\Lambda_{Yk}^*$ to an exponential (linear) cumulative hazard with rate equal to inverse sample mean of the waiting time. Throughout, we specify a fine partition with $\alpha_{Yk}$ close to zero. For small $\alpha_{Yk}$, realizations of the GP are widely dispersed around $\Lambda_{Yk}^*$ - representing a weak prior. Thus the posterior can deviate from this linear prior if the data suggests nonlinearity. An analogous model is specified for the waiting time to death after treatment $k$, $\lambda_{Tk}(w \mid  H_k,  \Lambda_{0Tk}, \beta_{Tk}) = \lambda_{0Tk}(w) \exp\big( H_k' \beta_{Tk} \big)$, with priors $\Lambda_{0Tk} \sim GP(\alpha_{Tk} \Lambda_{Tk}^*)$ and $\beta_{Tk} \sim f_{\beta_{Tk} }(\beta_{Tk} )$.

While the specification of confounder distribution is application specific, it is common to assume conditional independence among the $P$ covariates in $L_k$ at each time point such that $f_k(l_k \mid \bar H_{k-1} ) = \prod_{p=1}^P f_{kp}(l_{kp} \mid \bar H_{k-1} )$ and model each component separately. For generality of presentation, we write $f_k(l_k \mid \bar H_k, \eta_k )$, where $\eta_k$ is a vector of parameters that governs the joint distribution of the confounders at time $k$. These models can be as flexible or smooth as needed depending on sample size considerations. For instance, continuous $L_{kp}$ can be modeled as Gaussian with conditional mean dependent on $H_k$ through, say, a regression $E[L_{kp} \mid \bar H_k; \eta_{kp}] = H_k ' \eta_{kp}$, where $\eta_{kp}$ is a vector of regression coefficients that may include an intercept if a constant 1 is included in $H_k$. If more data is available we could be more flexible and specify $E[L_{kp} \mid \bar H_k; \eta_{kp}] = g(h_k; \eta_{kp})$, where $g$ is some function of history, and specify a BART \citep{Chipman2010} prior on the function  $g\sim BART$. Here, $\eta_{kp}$ would consist of the BART tree parameters. While time-varying covariates must usually be modeled as a function of history, nonparametric approaches like the Bayesian bootstrap can be used for the baseline confounder distribution, $f_1$. This is essentially a Bayesian analogue of the empirical distribution. Similarly, appropriate logistic models or BART models can be specified for binary covariates. Under the non-informative censoring assumption, all subjects who undergo course $k$ contribute to the likelihood for $\eta_k$ and to the hazards of the next possible events, $\lambda_{Tk}$ and $\lambda_{Yk}$ and time-varying covariate distributions. Details of the likelihood construction are given in the Supplement.

In all, this model is semi-parametric in the sense that we have a flexible specification for the waiting time hazards. It is generative in the sense that we model the full joint distribution of the data rather than, say, just the moments of some conditional outcome model. Generative models are useful because functionals of the joint can be computed easily by simulating from the model. For instance, in \eqref{eq:gcomp}, $\Psi^r(t)$ is just a functional of the unknown confounder distributions, $f_k$, and the unknown cause specific hazards, $\lambda_{Yk}$ and $\lambda_{Tk}$, which are completely characterized by the parameters $\omega = \{\omega_k \}_{k=1}^K$, where $\omega_k = \Big\{ \lambda_{0Tk}, \beta_{Tk}, \lambda_{0Yk},  \beta_{Yk}, \eta_k \Big\}$. Note that for the last course, $K$,  $\omega_K$ does not contain $\{  \lambda_{0YK},  \beta_{YK}\}$ since there is no subsequent course. Thus, from a Bayesian perspective, a posterior over $\omega$ induces a posterior over relevant functionals such as $\Psi^r(t)$, provided that we compute the requisite integrals.

\section{Adaptive Algorithm for Posterior Computation}
Inference for the required models follows from the joint posterior over the unknown cause-specific hazards and the parameters governing the sequence of confounder distributions, $ f_{\Omega | \mathcal {D} }( \omega_1, \omega_2, \dots, \omega_K \mid \mathcal{D} )$. In the following we will use bracket notation to denote generic posteriors. E.g. the joint posterior above will be denoted as $[ \omega_1, \omega_2, \dots, \omega_K ] $, where conditioning on $\mathcal{D}$ is suppressed for compactness. Since this posterior is not available in closed form, we implement a blocked Metropolis-in-Gibbs sampler. Specifically, let $ \omega_k^{(m)} = \Big\{ \lambda_{0Tk}^{(m)}, \beta_{Tk}^{(m)}, \lambda_{0Yk}^{(m)},  \beta_{Yk}^{(m)}, \eta_k^{(m)} \Big\}$ denote the $m^{th}$ posterior draw. We sample from the joint one block/variable at a time, conditional on the others, in the following way. First, we initialize values for all parameters to $\omega_k^{(0)}$ for all $k$ and. Then, at each iteration $m=1, 2, \dots, M$, update the model parameters for treatment courses $j=1, 2, \dots, K$ sequentially. Using the usual bracket notation to denotes posteriors, we begin by updating from conditional posterior of parameters at $j=1$, $\omega_1^{(m)} \sim  [\omega_1 \mid  \omega_{2}^{(m-1)}, \dots, \omega_{K}^{(m-1)}]$. To obtain $\omega_1^{(m)}= \{ \lambda_{0T1}^{(m)}, \beta_{T1}^{(m)}, \lambda_{0Y1}^{(m)},  \beta_{Y1}^{(m)}, \eta_1^{(m)} \} $, we update one component at a time conditional on the others. E.g. first, update $ \lambda_{0T1}^{(m)} \sim [  \lambda_{0T1} \mid \beta_{T1}^{(m-1)}, \lambda_{0Y1}^{(m-1)},  \beta_{Y1}^{(m-1)}, \eta_1^{(m-1)}]$, then update $\beta_{T1}^{(m)} \sim [\beta_{T1} \mid \lambda_{0T1}^{(m)} , \lambda_{0Y1}^{(m-1)},  \beta_{Y1}^{(m-1)}, \eta_1^{(m-1)} ]$, and so on. Having obtained $\omega_1^{(m)}$, we move to update $\omega_2^{(m)} \sim  [\omega_2 \mid  \omega_{1}^{(m)},  \omega_{2}^{(m-1)} , \dots, \omega_{K}^{(m-1)}]$ in the same way. Doing this for $j=1, 2, \dots, K$ yields the $m^{th}$ posterior draw $ \omega^{(m)}= \{\omega_1^{(m)}, \omega_2^{(m)}, \dots, \omega_K^{(m)}\} \sim [\omega_1, \omega_2, \dots, \omega_K]$. Typically we run the sampler for some amount of ``burn-in'' iterations $M^* < M$ and base inferences only on draws $M^* < m \leq M$.

Since conjugate priors for the coefficients of the proportional hazard model in \eqref{eq:gp} do not exist, we use a Metropolis step to update from their conditionals. The efficiency of the Metropolis algorithm greatly depends on the choice of proposal covariance, which can be difficult to tune manually in our setting given the number of parameters and models. Instead, we use an adaptive procedure \citep{Haario2001} that tunes the proposal covariance in the burn-in period. The sampler is initialized with a  symmetric, diagonal proposal covariance and run for $M^*$ iterations. At iteration $M^*$, we compute the estimated covariance of the $M^*$ draws. The remaining draws $M^* < m \leq M$ are then obtained via a proposal distribution with covariance equal to a scale factor of this estimated covariance. Details and forms for these conditionals are provided in the Supplement.

\subsection{Posterior G-computation and Maximization} \label{sc:gcomp}
In this section, we describe posterior g-computation and optimization for $\Psi^r(t)$ and utility functions more generally. In the Bayesian framework, we conduct g-computation for each posterior draw of the parameters $\{\omega_k^{(m)} \}_{k=1}^K$, for $m>M^*$. These parameters govern the continuous-time transition process between treatment and death as well as the time-varying confounder evolution over time. Given each draw of these parameters, the idea is to simulate from the process under hypothetical treatment rule $r$, then compute $\Psi^r(t)$ by averaging over those simulations. Here, we describe this process for $K=2$ for simplicity. Specifically, for the $m^{th}$ posterior draw, we do the following. For $b = 1, 2, \dots, B$
\begin{enumerate}
	\item Simulate confounder at first treatment course, $L_1^{(b)} \sim f_1(L_1 \mid \eta_1^{(m)})$ and make a treatment decision $ d_1^{r,(b)} = r_1(L_1^{(b)} )$
	\item Simulate time from first to second treatment: $W_{Y1}^{(b)} \sim \lambda_{Y1}(w \mid L_1^{(b)}, A_1 = d_1^{r,(b)}  \lambda_{0Y1}^{(m)}, \beta_{Y1}^{(m)} )$
	\item Simulate time from first treatment to death $W_{T1}^{(b)} \sim \lambda_{T1}(w \mid L_1^{(b)}, A_1 = d_1^{r,(b)},  \lambda_{0T1}^{(m)}, \beta_{T1}^{(m)} )$
	\item If $W_{T1}^{(b)} < W_{Y1}^{(b)}$, then set $T^{(b)} = W_{T1}^{(b)}$, $\kappa^{(b)} = 1$, and move to iteration $b+1$. Otherwise, store $W_{Y1}^{(b)}$, set $\kappa^{(b)} = 2$, and proceed to treatment course $k=2$:
	\item Simulate second course confounder $ L_2^{(b)} \sim f_2(L_2 \mid L_1^{(b)}, A_1 = d_1^{r,(b)}, W_{Y1}^{(b)},  \eta_2^{(m)}) $ and make a decision $ d_2^{r,(b)} = r_2( \bar L_2^{(b)}, A_1 = d_1^{r,(b)}, W_{Y1}^{(b)})  $
	\item Simulate $W_{T2}^{(b)} \sim \lambda_{T2}(w \mid \bar L_2^{(b)}, \bar A_2 = \bar d_2^{r,(b)},  W_{Y1}^{(b)}, \lambda_{0T2}^{(m)}, \beta_{T2}^{(m)} )$. Set $T^{(b)} =  W_{Y1}^{(b)} + W_{T2}^{(b)}$. 
	That is, total time is the waiting time between course one and two added with the waiting time from course two to death.
\end{enumerate}
After running for $B$ iterations, the $m^{th}$ posterior draw of the survival curve can be computed $\Psi^r(t)^{(m)} = B^{-1} \sum_{b=1}^B I( T^{(b)} > t )$.
The sampling of waiting times in steps 2,3, and 6 can be done efficiently via the inverse-CDF method under the Gamma Process models. For each $t$, this yields a set of $M-M^*$ posterior draws for $\Psi^r(t)$. We can form a point estimate by taking the mean across draws. Pointwise $100(1-\alpha)\%$ credible interval can be formed by using the $\alpha/2$ and $(1-\alpha/2)$ percentiles of draws as the endpoints. 

Comparing survival under two different rules $r$ and $r'$ does not require re-running the MCMC sampler. They are functionals of the same parameters, so we need only post-process the posterior draws, $\{\omega_k^{(m)} \}_{k=1}^K$, under another rule $r'$, as above, to produce draws for $\Psi^{r'}(t)$. In this way, we can also obtain posterior draws of contrasts such as $\Psi^{r}(t)/\Psi^{r'}(t)$ by dividing the posterior draws $\Psi^{r}(t)^{(m)}/ \Psi^{r'}(t)^{(m)}$. Absolute contrasts can be done by subtraction of draws. This is one of the main benefits of our generative approach. Once posterior draws of the nuisance parameters are obtained, they can be used to obtain posterior draws for complex transformations without re-running the sampler.

In addition to effect estimation, optimization within a class of rules $\mathcal{R}$ and can also be considered. For interpretability, we often restrict to simple rules that are characterized by a low-dimensional set of decision parameters $ \tau \in \Tau$, the class of rules is then given by $\mathcal{R} = \{ r_\tau : \tau \in \Tau \}$. In this case, we denote $r_\tau$ simply by $\tau$, since it completely determines the rule. In our setting, this is suitable since rules of interest are governed by a few thresholds for EF. The task is then to search $\Tau$ for the rule parameters that maximize the probability of surviving past some time, had we treated according to a rule with those parameters, $\Psi^{\tau}(t)$. Note that $\Psi^{\tau}(t) = \Psi^{\tau}(t; \omega )$ is a function of the unknown model parameters, $\omega$, written explicity as such for emphasis. Thus a posterior on these model parameters induces a posterior on the optimal rule parameters, $\tau^*( \omega) = \argmax_{\tau \in \mathcal{\Tau}} \Psi^{\tau}(t ; \omega)$. Computationally, maximization must be done for each posterior draw $\omega^{(m)}$, which yields a posterior draw of the optimal parameters $\tau^*( \omega^{(m)} ) \sim f( \tau^*( \omega) | \mathcal{D})$. In our application, $\tau$ is a pair of thresholds - one for relative change in EF and another for absolute EF - while $\Tau$ is specified as a two-dimensional grid of possible thresholds. This relatively small search space permits an exhaustive search over the grid. For each $\omega^{(m)}$, g-computation is done for each $\tau$ in the grid we choose $\tau^*( \omega^{(m)} )$ to be the threshold with the highest survival probability. This consititutes a draw from the bivariate posterior mass function (pmf), $f(\tau^*( \omega) | \mathcal{D})$. The posterior mode across the $m$ draws can be taken as a point estimate of $\tau^*$. A $100(1-\alpha)\%$ credible set of threshold pairs, $C(1-\alpha) \subset \Tau$, can be constructed by lowering a flat plane onto the pmf from above until $100(1-\alpha)\%$ of the mass lies above the plane. $C(1-\alpha)$ contains the pairs with mass above this plane. This is a bivariate analogue of the usual highest posterior density interval (HDI).

While our interest is primarily in overall survival, our generative approach allows for more general utility function outcomes \citep{Murray2017} that directly capture potential risk-benefit tradeoffs between survival and cardiotoxicity. For instance, consider $U^r(s, t) = \Psi^r(t) - \Phi^r(s)$, where $\Phi^r(s) = P( \min_{k>2} L_k(r) < s )$. If $L$ represents EF, then $\Phi^r(s)$ is the probability that EF drops below threshold $s$ after initial treatment. In the clinical literature, $s=.50$ is one common definition of an adverse cardiac event that we would ideally want to avoid. This function deducts utility based on the cardiac risk of the rule and can be computed just as above except at the end we compute $U^r( s, t )^{(m)} = \frac{1}{B} \sum_{b=1}^B  I( T^{(b)} > t) - I( L_2^{(b)} < s ) I(\kappa^{(b)}=2)$. We could consider more general rules that weigh the risk and benefit components differently. The idea is that rules can then be contrasted and optimized on the basis this utility. By construction, both the benefit ($\Psi^r(t)$) and the risk ($\Phi^r(s)$) are on the same (probability) scale, which eases interpretation. For instance, a rule where the risk exceeds benefit ($\Phi^r(s) >  \Psi^r(t)$) would have negative utility. If $r'$ has the same survival probability as $r$, but comes with lower risk, then it will have higher utility and would be preferred.

\section{Simulation Experiments}
Since this is a complex scenario with intricate treatment and time-varying confounding mechanisms, we ran extensive simulation studies evaluating the proposed method's ability to capture the survival time distribution in situations that mirror our application. Thus, the settings were carefully chosen to match our data: First, as in our data, there are four treatment courses. In a typical simulated data set, about $60\%$ of subjects complete the full sequence (the rest are either censored or die before completing) and about 40\% of the patients die while the remaining are censored. In the data, the corresponding proportions are about $60\%$ and $40\%$ as well. For each subject, we simulate two time-varying confounders - one continuous and one binary - as well as four time-constant confounders. This matches our data application which has a continuous and binary time-varying confounder (EF and presence of bloodstream infection, respectively) as well as several time-constant factors. At each course, we simulate treatment conditional on time-constant confounders, current and previous values of time-varying confounders, and previous treatment. Then, we simulate the waiting time until the next course, death, or censoring from separate Weibull proportional hazard models. Current and previous treatment, current and previous time-varying confounders, time-constant confounders, and the time since the previous course initiation all affect the waiting times until the next event by multiplying the Weibull baseline hazards. Longer times since last treatment are set to increase the time to next treatment, while decreasing time to death. This induces the positive correlation in Figure \ref{fig:descplots} and encodes the idea that longer waiting times between treatments indicate worse survival prospects. We simulate 1,000 such data sets in a large sample setting (n=1000) and small sample setting (n=300). For each simulated data set, we estimate posterior mean and 95\% credible intervals for $\Psi^r(t)$ at pre-specified $t$, with $r(L_{k1} ) = I( L_{k1} < 0)$. Here $L_{k1}$ is the continuous time-varying confounder with sufficient support around $0$, and $r$ assigns treatment $A_k=1$ if $L_{k1}<0$. The proposed approach is compared with a parametric version where the waiting times are modeled via a correctly specified Bayesian Weibull hazard model. The results are presented in Table \ref{tab:simres}. 

\begin{table}
\setlength{\extrarowheight}{-7pt}

\centering
\caption{Simulation Results: Absolute bias of posterior mean of $\Psi^r(t)$, for various $t$ (as percentage of the truth). Coverage of 95\% credible interval for $\Psi^r(t)$. Average width of 95\% credible interval.}
\label{tab:simres}
\resizebox{\linewidth}{!}{%
\begin{tabular}{l|l|rrr|rrr}
\multicolumn{1}{l}{}    & \multicolumn{1}{l}{}    & \multicolumn{3}{c}{Gamma Process}                                                             & \multicolumn{3}{c}{Weibull}                                                                    \\ 
\hline
Setting                 & $t$                      & \multicolumn{1}{l}{Bias (\%)} & \multicolumn{1}{l}{Coverage} & \multicolumn{1}{l}{Int. Width} & \multicolumn{1}{l}{Bias (\%)} & \multicolumn{1}{l}{Coverage} & \multicolumn{1}{l}{Int. Width}  \\ 
\hline
\multirow{4}{*}{N=1000} & \multicolumn{1}{c|}{5}  & 0.07                         & 95.4                         & 0.04                           & 0.14                          & 94.5                         & 0.04                            \\
                        & \multicolumn{1}{c|}{10} 		  & 0.03                          & 95.2                         & 0.07                           & 0.22                          & 94.8                         & 0.07                            \\
                        & \multicolumn{1}{c|}{15} 		  & 2.29                          & 90.4                         & 0.08                           & 0.24                          & 95.8                         & 0.08                            \\
                        & \multicolumn{1}{c|}{20} 		  & 185.00                      & 18.8                         & 0.13                           & 6.58                          & 94.5                         & 0.06                            \\ 
\hline
\multirow{4}{*}{N=300}   & 5                       	         & 0.34                            & 94.9                         & 0.07                           &    0.53                          & 95.8                         & 0.07                            \\
                  		     & 10                              & 0.13                          & 94.7                         & 0.12                           &    0.98                          & 95.2                         & 0.12                            \\
                        		     & 15                              & 4.52                            & 89.3                         & 0.14                           &    0.51                          & 94.9                         & 0.14                            \\
                        		     & 20                               & 227.00                        & 14.7                         & 0.18                           & 25.00                          & 90.3                         & 0.11                            \\
\hline
\end{tabular}
}
\end{table}

Across simulations, the proportion of patients still at risk for subsequent treatment or death at time points 5, 10, 15, and 20 are about 95\%, 85\%, 60\%, and 2\%, respectively. Accordingly, at early time points, $t \in \{5, 10, 15\}$ the bias is quite low: .07\%, .03\%, and 2.29\%, respectively, in the $N=1000$ setting. Similarly, the coverage of the 95\% credible intervals are close to nominal. At $t=20$, only  about 2\% of subjects are left in the sample in a typical simulation run. Thus, the quality of the semiparametric estimate is degraded. With such low proportion at risk, all flexible estimators will perform poorly at the tails. The Weibull model is correctly specified in this setting, so it does better in terms of bias and coverage even at the tail. In practical applications, however, a Weibull model is unlikely to be correct in applications and we would avoid extrapolating so far into the tail. These results demonstrate that GP-based posterior estimates have comparable frequentist properties to the correctly specified model in a setting with complex time-varying confounding and informative treatment timing. Despite being unspecified, the GP posterior picks up the Weibull baseline hazard form. The small sample setting with $N=300$ is more consistent with our data application and exhibits largely the same performance. As expected, interval widths tend to be larger for both models relative to the larger sample case. Similarly, bias tends to be higher relative to the large sample setting as well. Additional details on model specification and other settings can be found in the Supplement. Because the GP models require a partitioning of the time interval, the appendix includes a sensitivity for the $n=1000$ setting where the number of partitions is halved, yielding a coarser partition. The results are generally consistent with only a slight increase in bias and coverage (1\% and 96\%, respectively)

\section{Evaluating Ejection-Fraction Based Treatment Rules}
We analyze data from the AAML1031-PHIS cohort with n=292 subjects who had a median followup time of 2.5 years and underwent a maximum of four ACT treatment courses. While a small sample, there are only about 500 new diagnoses of pediatric AML annually in the United States. Table \ref{tab:sumstats} displays some sample statistics for our data. About 36\% of the patients either died or were censored before completing the four-course treatment protocol. Death is observed for about 40\% of patients. As illustrated in Figure \ref{fig:descplots}, the distributions of the waiting times between treatments tend to be centered around 35 days, with some right-skewness due to subjects who take longer than usual to recover from their previous course. This motivates the need for flexible models for these waiting times.

\begin{table}[h!]
\centering
\caption{Summary Statistics: Median and interquartile range presented for continuous features. Counts and proportions presented for discrete features.  \label{tab:sumstats}}
\begin{tabular}{l|l} 
\multicolumn{1}{c}{Features}      & \multicolumn{1}{c}{ $N=292$ } \\
\hline
Follow-up (years)                                   		& 2.6 (1.3-3.7 )          \\
Death                                               		& 114 (40\%)  		\\
Num. Treatment Courses, ($\kappa$) 		&                             		\\
\multicolumn{1}{c|}{1}                              		& 22 (8\%)    		\\
\multicolumn{1}{c|}{2}                              		& 36 (12\%)  		 \\
\multicolumn{1}{c|}{3}                              		& 46 (16\%)    		 \\
\multicolumn{1}{c|}{4}                              		& 188 (63\%)  		 \\
AML Risk Classification (high)                                        		& 64 (23\%)  		 \\
WBC (cells/uL)                                                 			& 20 (6.8-65)              \\
Male                                                			& 152 (52\%)  		 \\
\hline
\end{tabular}
\end{table}

We also observe several important patient-level confounders. Time-varying confounders include EF (recorded ahead of each treatment decision) and the (binary) presence/absence of a bloodstream infection since the start of the previous treatment course, which we denote as $V_k$. The treatment options of interest consist of chemotherapy treatments that include ($A_k=1$) or exclude ACT ($A_k=0$), which is recorded for each visit. Each of the two possible treatments are feasible in courses 1,2, and 4 for all histories. That is $\mathcal{S}_k(h_k) = \{0,1\}$ for all $h_k$ at $k=1,2, 4$. However, in general subjects were not eligible for ACT at course three, meaning $\mathcal{S}_3(h_3) = \{0\}$, $\forall h_3$. Several potential baseline confounders including age, sex, baseline white blood cell (WBC) count, and AML risk classification are also recorded. Baseline white blood cell count indicates amount of leukemia at the start of the treatment protocol and may inform subsequent ACT decisions and survival. AML risk classification (RS) is a binary covariate (high versus low) based on favorable/unfavorable molecular and cytogenetic features linked with prognosis of worse survival. We assume sequential ignorability and non-informative censoring (as well as the other identification assumptions discussed) all hold conditional on these factors. While these assumptions cannot be verified in the data, we know that these covariates (EF in particular) play an influential role in treatment decisions.

Using the notation of our model, $L_k = (EF_k, V_k, age, sex, WBC, RS)$. For a subject reaching course $k$, the available history is $H_k = ( \bar A_{k-1}, \bar W_{k-1}, \bar{EF}_k, \bar{V}_k, age, sex, WBC, RS)$. We fit the models described in Section \ref{sc:models}, adjusting for information in current and previous treatment course in all models. Since $EF_k \in(0,1)$ is a proportion, at each course we model its distribution as a Beta with conditional mean being a function of history from course $k-1$ through a logit link. Similarly, we model $V_k$ with a sequence of logistic regressions to respect the binary domain. At each course, the hazard models for subsequent death or treatment conditions on current EF, $V$, and ACT decisions as well as their values from the previous treatment course in addition to all baseline confounder. To avoid linearity assumptions, the time since previous treatment, $W_{k-1}$, is included as a categorical variable with four levels with cutoffs at the 25th, 50th, and 75th percentiles of the observed values. Detailed specifications of these models are provided in the Supplement. We use the outlined MCMC approach to obtain draws from the posterior of all the unknowns.

\begin{figure}
\begin{center}
\includegraphics[width=1\linewidth]{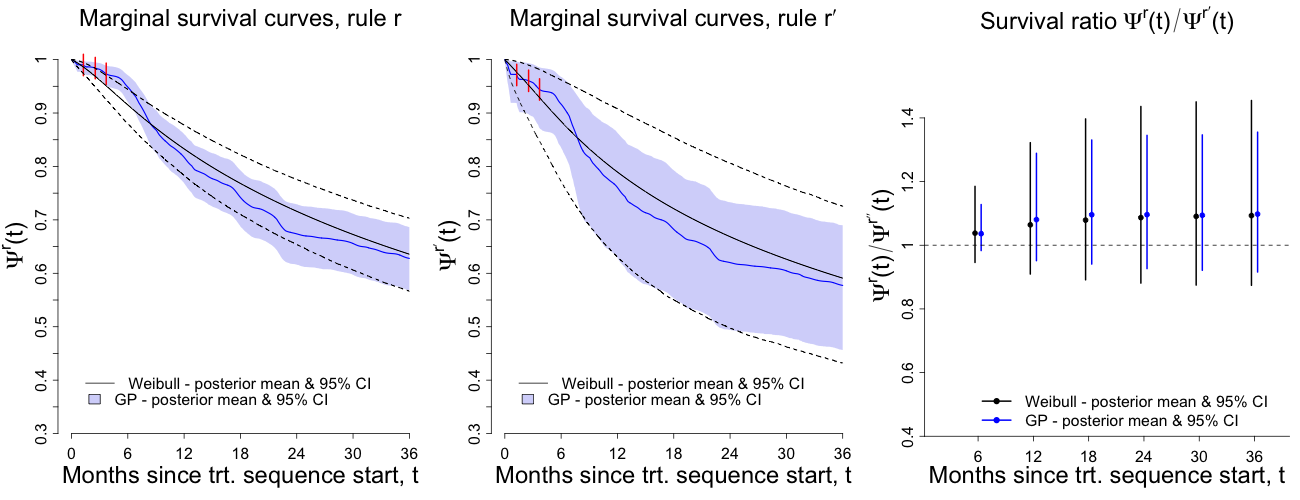}
\end{center}
\caption{ Left: Posterior inference for $\Psi^r(t)$ under GP versus Weibull models for $r$. Red ticks indicate median time of treatment sequence 2, 3, and 4 initiation in order. Middle: Posterior inference for $\Psi^{r'}(t)$ under GP versus Weibull models for $r$. Right: risk ratios comparing rule $r$ with $r'$. \label{fig:survres} }
\end{figure}

Current criteria for ACT withholding varies between trials and thresholds have not been rigorously validated. For instance, one rule that has been recommended in the clinical literature is to withhold ACT ($A_k=0$) if current EF has declined $10\%$ from baseline to below $50\%$. I.e. $A_k = r_k( EF_k, EF_1 )$, where $ r_k( EF_k, EF_1 ) = 1 - I( EF_k/EF_1 - 1 < -.1 )I( EF_k < .5 )$ \citep{Neuendorff2020}. To assess the impact of this rule, the left panel of Figure \ref{fig:survres} visualizes posterior inference for survival, $\Psi^r(t)$, under rule $r( \bar EF_K ) =  ( A_1 = 1 - I( EF_1 < .5 ), A_2= r_2( EF_2, EF_1 ), A_3=0, A_4 = r_4( EF_4, EF_1 ) )$.  At $k=3$, the rule always withholds ACT since it is not feasible at the third course. Figure \ref{fig:survres} presents results under both the outlined GP model and a parametric Weibull model. Note that the first treatment is assigned only using baseline EF since percentage change from itself is zero. It can be seen from Figure \ref{fig:survres} that the parametric model smooths over intricacies of the survival function, whereas the GP captures important features - for instance, the steep drop-off in survival around month 4. In the middle panel, we present the posterior survival curve for rule $r'( \bar EF_K ) =  ( A_1 = 0, A_2= 1 - I( EF_2/EF_1 - 1 < -.1 ), A_3=0, A_4 = 1 - I( EF_4/EF_1 - 1 < -.1 ) )$. This starts everyone on nACT, then withholds ACT only if relative change in EF decreases more than 10\% (regardless of absolute EF level). This rule relaxes the criterion for withholding ACT  begins by initially withholding ACT. The right panel presents the ratio in survival probabilities for rule $r$ versus $r'$ and shows relative survival benefit for $r$, which has stricter criteria for withholding ACT. In summary, posterior mean 3-year survival probability under $r$ is $.64$ $[95\% \text{CrI}: .58 - .69]$. Under rule $r'$ posterior mean 3-year survival probability is $ .60$ $[95\% \text{CrI}: .46 - .7]$. The right panel of  Figure \ref{fig:survres} shows the relative survival probabilities under each rule at various time points. There is higher survival under rule $r$ across time.

\begin{figure}
\begin{center}
\includegraphics[width=6in]{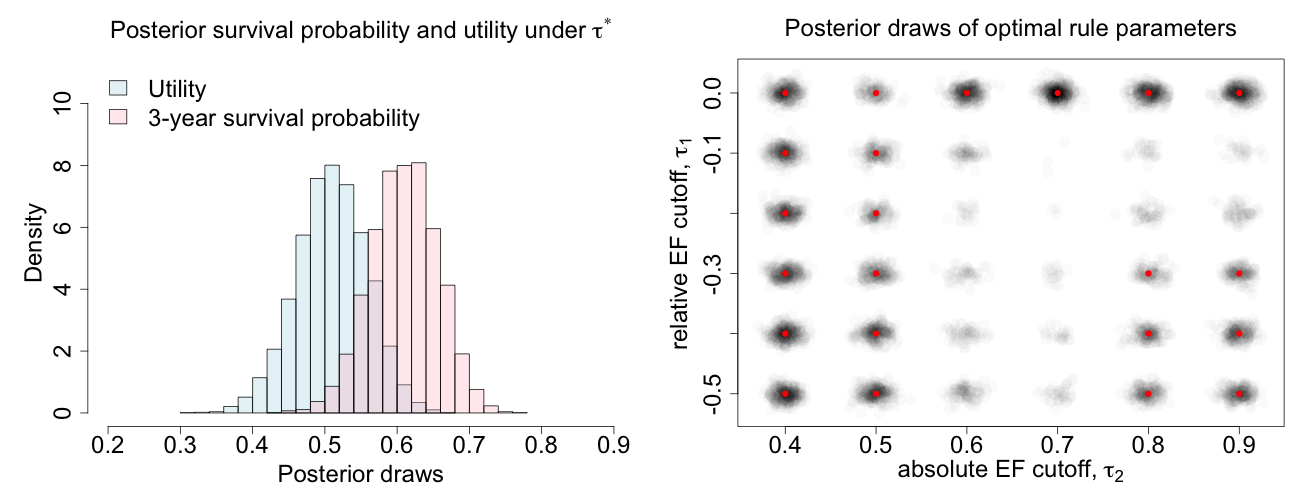}
\end{center}
\caption{ Left: Posterior distribution of 3-year survival and utility under posterior mode optimal rule, $(\tau_1=0, \tau_2=.7)$. Under this rule, posterior mean 3-year survival is about 60\% with about 10\% risk of EF dipping below .5 during the treatment course. Right: Posterior draws of the optimal rule $\tau^*$. Red points indicate threshold pairs in the 90\% credible set, $C(1-\alpha) \subset \Tau$ for $\alpha=.10$ \label{fig:optim}}
\end{figure}

The actual thresholds of $.5$ and $ -.1$ in $r$ are still debated in the literature. To inform this debate, we consider finding thresholds that optimize 3-year survival, an important endpoint in pediatric AML. Specifically, we define a class of rules $r(\bar EF_K; \tau) = ( A_1 = 1 - I( EF_1 < \tau_2 ), A_2=r_2 ( EF_2, EF_1; \tau ) ,  A_3 = 0 , A_4 = r_k ( EF_4, EF_1; \tau ) )$. Here, $r_k( EF_k, EF_1; \tau ) = 1 - I( EF_k/EF_1 - 1 < \tau_1 )I( EF_k < \tau_2 )$ and $\tau = (\tau_1, \tau_2)$ parameterizes the rule. For $ \tau\in \mathcal{\Tau} =  \{ 0, -.1, -.2, -.3, -.4, -.5 \}  \times \{ .4, .5, .6, .7, .8, .9 \}$, this defines a class of rules, where the range of threshold values are selected with clinical guidance and positivity constraints in mind. We optimize this rule by finding a posterior over $\tau^* = \argmax_{\tau \in \mathcal{\Tau} } U^\tau (36, .5) $, where $ U^\tau (36, .5)$ is the 3-year utility had we treated according to rule $r(\bar EF_K; \tau)$. As discussed in Section \ref{sc:gcomp}, this utility is the risk of post-baseline EF dipping below .5 subtracted from the 3-year survival probability. We do this as outlined in Section \ref{sc:gcomp} and display the posterior draws of $\tau^*$ as points in the right panel over Figure \ref{fig:optim}. The points are jittered and transparent to create a kind of heat map with more posterior draws leading to darker clouds at that $\tau$ pair. The posterior mode threshold pair is $(\tau_1 = 0, \tau_2 = .7)$. This favors a rule that withholds ACT if there is any decline from baseline, as long as EF falls below .7. If a patient experiences an increase in EF from baseline, then ACT is assigned. This makes clinical sense as ACT is cardiotoxic and we may wish to assign it if heart function is improved. The red points in the figure indicate all threshold pairs that are in the 90\% credible set, $C(.90) \subset \Tau$, as discussed in Section \ref{sc:gcomp}. Note that the previously discussed rule of thumb with $(\tau_1 = -.1, \tau_2 = .5 )$ is in this 90\% credible set and, in that sense, is justified by this data. The left panel of Figure \ref{fig:optim} are draws from the posterior of the 3-year survival probability under posterior mode rule, $(\tau_1 = 0, \tau_2 = .7)$. The posterior distribution of the utility is shifted left since there is about a 10\% posterior mean risk of EF dipping below .5 under this rule. 

This approach characterizes the optimal rule as an unknown about which we have uncertainty. There are many ways this can be translated to inform clinical decision-making in practice. One possibility is outputting all rules in the credible set $C(.90)$ and leaving the final decision up to the clinician. The credible set gives some guidance about how well-supported various rules are by the data. Another option is to provide the entire list of thresholds, sorted by the posterior probability of each threshold pair being the optimal. The clinician may be wary of choosing pairs at the bottom of the list and more comfortable choosing ones higher on the list. In either case, we view this output as a decision support tool that allows clinicians to gauge how strongly the data support various treatment options. The ultimate decision is made by the clinician and patient together.

\addtolength{\textheight}{.5in}%

\section{Discussion}
\label{sec:disc}

Approaches for estimating causal effects of dynamic treatment rules exist, but our data presented several unique challenges: a fixed number of treatment courses that occur at random, informative times and occurrence of death and censoring that prevent subjects from completing the full sequence of treatment courses. We use results from causal inference to identify a causal survival probability under a particular dynamic treatment rule and specify a generative Bayesian semiparametric model for how patients transition from one treatment to subsequent treatment (or death) conditional on treatment and confounder history. An appealing aspect of this approach is full posterior inference for all functions of the transition process - including causal survival probabilities, contrasts of survival probabilities under alternate rules, and even posterior inference over optimal rules. In simulations we assess the frequentist performance of this Bayesian estimator and report nominal coverage and low bias for the 95\% credible interval and posterior mean survival probability, respectively. In our analysis of data from the AAML trial, we show that our semiparametric model captures complexities in the survival curve that the parametric Weibull model smooths over. We also illustrate that current rules of thumb for ACT administration fall within a 90\% credible set of optimal rules - indicating that current practice has some support in the data. 

A central difficulty with time-varying treatments is the difficulty in estimating the effect of all possible treatment sequences in small samples. Suppose there are four treatment courses with two possible treatments in each course leading to $2^4$ possible treatment sequences. In small samples there may be very few patients who follow some of these sequences, essentially leading to random violations of positivity. An appealing aspect of this Bayesian approach is that the credible intervals on the estimated survival probabilities under such sequences are quite wide - reflecting our increased uncertainty due to sparsity of data. For instance, when we try to estimate the effect of a rule that may assign ACT in course three, the 95\% credible interval around 3-year survival probability covers almost the entire range $[0,1]$. Due to the fact that ACT is not feasible (only two subjects were assigned ACT in the third course as exceptions to protocol), our credible intervals widen to reflect the increased uncertainty that stems from extrapolating beyond the range of the observed data.

\subsection*{Acknowledgements and Funding}
This work was partially funded by: 1) the Patient-Centered Outcomes Research Institute (PCORI) contract ME-2021C3-24942; 2) National Institutes of Health K Award K01HL143153; 3); National Institutes of Health R01 Award R01HL164925. 4) Brown University Office of the Vice President for Research (OVPR Salomon Faculty Research Award awarded.

\bibliographystyle{chicago}
\bibliography{main.bib}

\begin{thebibliography}{}

\bibitem[\protect\citeauthoryear{Chipman, George, and McCulloch}{Chipman
  et~al.}{2010}]{Chipman2010}
Chipman, H.~A., E.~I. George, and R.~E. McCulloch (2010).
\newblock {BART: Bayesian additive regression trees}.
\newblock {\em The Annals of Applied Statistics\/}~{\em 4\/}(1), 266--298.

\bibitem[\protect\citeauthoryear{Getz, Sung, Ky, Gerbing, Leger, Leahy, Sack,
  Woods, Alonzo, Gamis, and Aplenc}{Getz et~al.}{2019a}]{Getz2019}
Getz, K.~D., L.~Sung, B.~Ky, R.~B. Gerbing, K.~J. Leger, A.~B. Leahy, L.~Sack,
  W.~G. Woods, T.~Alonzo, A.~Gamis, and R.~Aplenc (2019a).
\newblock Occurrence of treatment-related cardiotoxicity and its impact on
  outcomes among children treated in the aaml0531 clinical trial: A report from
  the children’s oncology group.
\newblock {\em Journal of Clinical Oncology\/}~{\em 37\/}(1), 12--21.
\newblock PMID: 30379624.

\bibitem[\protect\citeauthoryear{Getz, Sung, Ky, Gerbing, Leger, Leahy, Sack,
  Woods, Alonzo, Gamis, and Aplenc}{Getz et~al.}{2019b}]{Getz2019_2}
Getz, K.~D., L.~Sung, B.~Ky, R.~B. Gerbing, K.~J. Leger, A.~B. Leahy, L.~Sack,
  W.~G. Woods, T.~Alonzo, A.~Gamis, and R.~Aplenc (2019b).
\newblock Occurrence of treatment-related cardiotoxicity and its impact on
  outcomes among children treated in the aaml0531 clinical trial: A report from
  the children’s oncology group.
\newblock {\em Journal of Clinical Oncology\/}~{\em 37\/}(1), 12--21.
\newblock PMID: 30379624.

\bibitem[\protect\citeauthoryear{Guan, Reich, Laber, and Bandyopadhyay}{Guan
  et~al.}{2020}]{Qian2020}
Guan, Q., B.~J. Reich, E.~B. Laber, and D.~Bandyopadhyay (2020).
\newblock Bayesian nonparametric policy search with application to periodontal
  recall intervals.
\newblock {\em Journal of the American Statistical Association\/}~{\em
  115\/}(531), 1066--1078.
\newblock PMID: 33012901.

\bibitem[\protect\citeauthoryear{Haario, Saksman, and Tamminen}{Haario
  et~al.}{2001}]{Haario2001}
Haario, H., E.~Saksman, and J.~Tamminen (2001).
\newblock {An adaptive Metropolis algorithm}.
\newblock {\em Bernoulli\/}~{\em 7\/}(2), 223 -- 242.

\bibitem[\protect\citeauthoryear{Hua, Mei, Zohar, Giral, and Xu}{Hua
  et~al.}{2021}]{Hua2021}
Hua, W., H.~Mei, S.~Zohar, M.~Giral, and Y.~Xu (2021).
\newblock {Personalized Dynamic Treatment Regimes in Continuous Time: A
  Bayesian Approach for Optimizing Clinical Decisions with Timing}.
\newblock {\em Bayesian Analysis\/}, 1 -- 30.

\bibitem[\protect\citeauthoryear{Kalbfleisch}{Kalbfleisch}{1978}]{Kalbfleisch1978}
Kalbfleisch, J.~D. (1978).
\newblock Non-parametric bayesian analysis of survival time data.
\newblock {\em Journal of the Royal Statistical Society. Series B
  (Methodological)\/}~{\em 40\/}(2), 214--221.

\bibitem[\protect\citeauthoryear{Kalbfleisch and Prentice}{Kalbfleisch and
  Prentice}{2011}]{kalbfleisch2011}
Kalbfleisch, J.~D. and R.~L. Prentice (2011).
\newblock {\em The statistical analysis of failure time data}.
\newblock John Wiley \& Sons.

\bibitem[\protect\citeauthoryear{Murphy}{Murphy}{2003}]{Murphy2003}
Murphy, S.~A. (2003).
\newblock Optimal dynamic treatment regimes.
\newblock {\em Journal of the Royal Statistical Society. Series B (Statistical
  Methodology)\/}~{\em 65\/}(2), 331--366.

\bibitem[\protect\citeauthoryear{Murray, Thall, Yuan, McAvoy, and Gomez}{Murray
  et~al.}{2017}]{Murray2017}
Murray, T.~A., P.~F. Thall, Y.~Yuan, S.~McAvoy, and D.~R. Gomez (2017).
\newblock Robust treatment comparison based on utilities of semi-competing
  risks in non-small-cell lung cancer.
\newblock {\em Journal of the American Statistical Association\/}~{\em
  112\/}(517), 11--23.
\newblock PMID: 28943681.

\bibitem[\protect\citeauthoryear{Neuendorff, Loh, Mims, Christofyllakis, Soo,
  Bölükbasi, Oñoro-Algar, Hundley, and Klepin}{Neuendorff
  et~al.}{2020}]{Neuendorff2020}
Neuendorff, N.~R., K.~P. Loh, A.~S. Mims, K.~Christofyllakis, W.-K. Soo,
  B.~Bölükbasi, C.~Oñoro-Algar, W.~G. Hundley, and H.~D. Klepin (2020, 02).
\newblock {Anthracycline-related cardiotoxicity in older patients with acute
  myeloid leukemia: a Young SIOG review paper}.
\newblock {\em Blood Advances\/}~{\em 4\/}(4), 762--775.

\bibitem[\protect\citeauthoryear{Robins}{Robins}{1986}]{Robins1986}
Robins, J. (1986).
\newblock A new approach to causal inference in mortality studies with a
  sustained exposure period—application to control of the healthy worker
  survivor effect.
\newblock {\em Mathematical Modelling\/}~{\em 7\/}(9), 1393--1512.

\bibitem[\protect\citeauthoryear{Tsiatis, Davidian, Holloway, and
  Laber}{Tsiatis et~al.}{2019}]{tsiatis2019}
Tsiatis, A.~A., M.~Davidian, S.~T. Holloway, and E.~B. Laber (2019).
\newblock {\em Dynamic treatment regimes: Statistical methods for precision
  medicine}.
\newblock Chapman and Hall/CRC.

\bibitem[\protect\citeauthoryear{Xu, Müller, Wahed, and Thall}{Xu
  et~al.}{2016}]{Xu2016}
Xu, Y., P.~Müller, A.~S. Wahed, and P.~F. Thall (2016).
\newblock Bayesian nonparametric estimation for dynamic treatment regimes with
  sequential transition times.
\newblock {\em Journal of the American Statistical Association\/}~{\em
  111\/}(515), 921--950.
\newblock PMID: 28018015.

\end{thebibliography}

\end{document}